\documentclass{article}

\usepackage{authblk}
\usepackage{ulem}
\usepackage{caption}
\usepackage{subcaption}
\usepackage{float}
\usepackage{longtable}
\usepackage{tabularray}
\usepackage{wrapfig}
\usepackage[mode=text]{siunitx}
\usepackage{graphicx}
\usepackage{natbib}
\usepackage{tabularx}

\usepackage{listings}
\usepackage{xcolor}

\definecolor{codegreen}{rgb}{0,0.6,0}
\definecolor{codegray}{rgb}{0.5,0.5,0.5}
\definecolor{codepurple}{rgb}{0.58,0,0.82}
\definecolor{backcolour}{rgb}{0.95,0.95,0.92}

\lstdefinestyle{pythonstyle}{
    backgroundcolor=\color{backcolour},   
    commentstyle=\color{codegreen},
    keywordstyle=\color{magenta},
    numberstyle=\tiny\color{codegray},
    stringstyle=\color{codepurple},
    basicstyle=\ttfamily\footnotesize,
    breakatwhitespace=false,         
    breaklines=true,                 
    captionpos=b,                    
    keepspaces=true,                 
    numbers=left,                    
    numbersep=5pt,                  
    showspaces=false,                
    showstringspaces=false,
    showtabs=false,                  
    tabsize=2,
    frame=single,
    rulecolor=\color{black}
}

\title{Leveraging Modified \textit{Ex Situ} Tomography Data for Segmentation of \textit{In Situ} Synchrotron X-Ray Computed Tomography}

\author[1,2]{Tristan Manchester}
\author[1,2]{Adam Anders}
\author[1,2]{Julio Spadotto}
\author[1,2]{Hannah Eccleston}
\author[1,2]{William Beavan}
\author[3]{Hugues Arcis}
\author[1,2]{Brian J.~Connolly}

\affil[1]{Department of Materials, University of Manchester, Manchester, M13 9PL, United Kingdom}
\affil[2]{Henry Royce Institute, Royce Hub Building, Oxford Rd, Manchester, M13 9PL, United Kingdom}
\affil[3]{United Kingdom National Nuclear Laboratory, Building D5, First Floor, Culham Campus, Abingdon, OX14 3DB, United Kingdom}

\date{Correspondence: tmanchester96@gmail.com}

\begin{document}

\maketitle

\begin{abstract}
\textit{In situ} synchrotron X-ray computed tomography enables dynamic material studies. However, automated segmentation remains challenging due to complex imaging artefacts---like ring and cupping effects---and limited training data. We present a methodology for deep learning-based segmentation by transforming high-quality \textit{ex situ} laboratory data to train models for segmentation of \textit{in situ} synchrotron data, demonstrated through a metal oxide dissolution study. Using a modified SegFormer architecture, our approach achieves segmentation performance (94.7\% IoU) that matches human inter-annotator reliability (94.6\% IoU). This indicates the model has reached the practical upper bound for this task, while reducing processing time by 2 orders of magnitude per 3D dataset compared to manual segmentation. The method maintains robust performance over significant morphological changes during experiments, despite training only on static specimens. This methodology can be readily applied to diverse materials systems, enabling the efficient analysis of the large volumes of time-resolved tomographic data generated in typical \textit{in situ} experiments across scientific disciplines.

\end{abstract}

\section{Introduction}
The ability to non-destructively visualise and quantify the internal structure and evolution of materials has revolutionised numerous fields, from materials science and geosciences to biomedical engineering and energy storage technologies. One such powerful technique that has enabled this transformative capability is X-ray computed tomography (XCT), which allows for the three-dimensional reconstruction of a sample's internal microstructure through the acquisition and computational processing of a series of two-dimensional radiographs. This technique provides researchers with comprehensive volumetric data that reveals features and characteristics that would otherwise remain hidden within the material's structure.

Synchrotron XCT has emerged as an invaluable tool for investigating dynamic material processes \cite{withersXrayComputedTomography2021}. In contrast to conventional laboratory-based X-ray sources, the exceptional brilliance of synchrotron radiation means complete tomographic datasets can be acquired within seconds to minutes, rather than hours, enabling significantly greater temporal resolution. This temporal advantage is crucial for \textit{in situ} experiments of changing systems, like metal oxide dissolution, where morphological changes can occur on timescales shorter than the several hours required for laboratory XCT acquisition. Such extended scan times would not only fail to capture intermediate dissolution states, but also introduce motion-induced blurring as the specimen continues to evolve during data collection\cite{fineganInoperandoHighspeedTomography2015,loweMetamorphosisRevealedTimelapse2013,vanoffenwertPoreScaleVisualizationQuantification2019}. However, synchrotron imaging presents distinct challenges. Reconstructed tomograms often exhibit artefacts such as rings, beam-hardening-induced cupping effects, motion-induced blurring, and reduced signal-to-noise ratios (SNR), stemming from a variety of sources \cite{chao-kungyangSimulationStudyMotion1982,toftsSourcesArtefactComputed1980,ohnesorgeEfficientCorrectionCT2000,rabrooksBeamHardeningXray1976,kakPrinciplesComputerizedTomographic2001}. 
\begin{figure}[h]
      \centering
      \includegraphics[width=\textwidth]{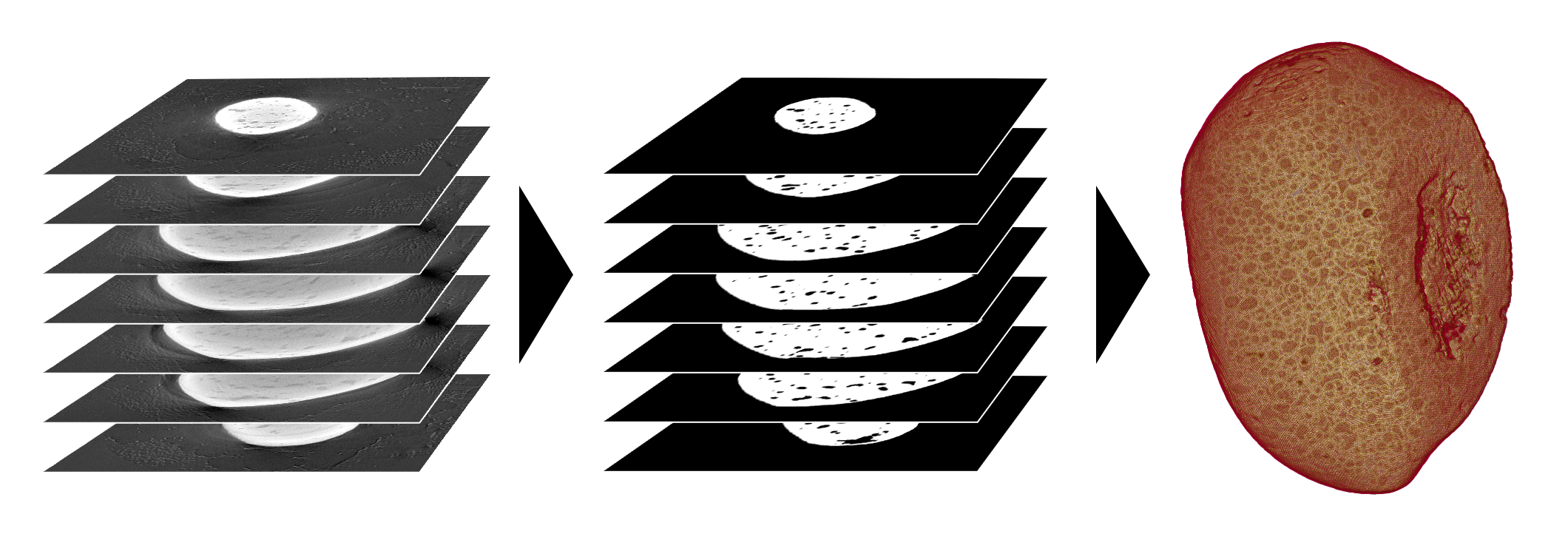}
\caption{Illustration of X-ray computed tomography segmentation. Reconstructed 2D XCT reconstruction slices (left) are segmented to differentiate phases or features (middle), allowing for the reconstruction of a 3D volume (right) for subsequent quantification and analysis.}
      \label{fig:segmentation_concept}
\end{figure}
Although qualitative analysis of X-ray tomograms is valuable, the true analytical power of XCT emerges when images are quantitatively assessed through a process known as segmentation \cite{withersXrayComputedTomography2021}. Segmentation assigns each voxel (3D pixel) in the volume to a particular phase or material within the specimen, enabling the extraction of critical metrics such as porosity, volume fraction, and tortuosity, as illustrated in Figure~\ref{fig:segmentation_concept}. The simplest segmentation methods are threshold-based approaches, whereby phases that exhibit different X-ray attenuation coefficients present with distinct grey values in the final reconstruction image, with clear boundaries between phases \cite{sheppardTechniquesImageEnhancement2004,torralbaComparisonSurfaceExtraction2018}. This allows these voxels to be reliably assigned a particular label. Beyond thresholding, researchers have employed more sophisticated techniques, including region-growing algorithms, watershed segmentation, and clustering methods such as k-means \cite{sheppardTechniquesImageEnhancement2004,torralbaComparisonSurfaceExtraction2018,borgesdeoliveiraExperimentalInvestigationSurface2016,sokacImprovedSurfaceExtraction2020}. However, the aforementioned artefacts can significantly complicate this process, rendering the boundaries between different materials ambiguous to the algorithms. Manual segmentation offers an obvious alternative: human operators can more readily discern the true structure of the data and label it by hand. This approach, however, is prohibitively time-consuming even for a single tomogram containing hundreds to thousands of 2D slices; with time-resolved studies comprising tens or hundreds of tomograms, manual segmentation becomes entirely impractical, creating a significant bottleneck in the analysis workflow.

Deep learning---a form of machine learning via neural networks---has emerged as a compelling alternative to conventional segmentation methods, most notably since the introduction of the U-net neural network architecture for medical image segmentation \cite{ronnebergerUNetConvolutionalNetworks2015}. The application of such networks to synchrotron data has rapidly advanced the state-of-the-art for analysing complex microstructures and imaging artefacts. For instance, U-Net architectures have been adapted to segment multiphase metallic alloys, with performance enhanced by implementing pixel-wise weighted loss functions that prioritise the identification of critical but hard-to-detect features \cite{strohmannSemanticSegmentationSynchrotron2019}. Other models like Sensor3D have been used to overcome the non-linear halo and shade-off artefacts inherent to specific modalities like Zernike phase-contrast nano-tomography, enabling the automated analysis of bone porosity \cite{silveiraDeepLearningOvercome2024}. The high data rates of \textit{in situ} experiments have spurred the development of bespoke, lightweight networks such as AM-SegNet, designed for rapid segmentation and quantification of radiographic data from dynamic processes like additive manufacturing \cite{liAMSegNetAdditiveManufacturing2024}. Beyond direct segmentation of reconstructed volumes, more sophisticated strategies have been proposed, such as multi-stage pipelines that apply separate deep learning models to projections, sinograms, and reconstructions, targeting artefacts in their respective domains where they are most effectively addressed \cite{shiMultistageDeepLearning2025}. To make these tools more accessible, user-focused platforms like MLExchange are also being developed, often incorporating efficient architectures such as Mixed-Scale Dense Convolutional Neural Networks (MSDNet) to handle large data volumes at scientific user facilities \cite{haoDeployingMachineLearning2023}.

Neural networks are trained by providing them with exemplary data that has already been accurately segmented to establish a "ground truth". The network essentially generates predictions and is subsequently assigned a score indicating the accuracy of its performance during each iteration, known as the loss function. The objective is to minimise this loss function, achieved via gradient descent optimisation, whereby the network gradually learns to replicate the segmentation process. This process typically requires a substantial volume of training data---often thousands of examples---and herein lies the paradox: neural networks are required to segment large datasets, yet large segmented datasets are required to train neural networks. As noted previously, manual segmentation is a time-consuming and tedious process, and inherently subjective, meaning different researchers assigned to this arduous task might produce varying annotation results for challenging datasets.
While architectures like generative adversarial networks (GANs), which employ a pair of neural networks working in opposition to one another, have been explored to create synthetic training data in fields such as medical imaging \cite{usmanakbarBrainTumorSegmentation2024,thambawitaSinGANSegSyntheticTraining2022, koetzierGeneratingSyntheticData2024,fengEnhancingMedicalImaging2024} and materials science \cite{sardharaGenerativeAdversarialNetworks2025}, they present their own challenges. These include challenges in ensuring sufficient realism and diversity in generated data, potential performance gaps compared to authentic training, higher computational overhead for high-resolution data, and training instability \cite{huDiscriminatorCooperatedFeatureMap2023,wangQGANQuantizedGenerative2019,saxenaGenerativeAdversarialNetworks2022,chenChallengesCorrespondingSolutions2021,manishaGenerativeAdversarialNetworks2018}.

Unsupervised learning approaches have also garnered attention as potential solutions to the annotation burden. Self-supervised methods, which learn representations from unlabelled data through pretext tasks such as rotation prediction or image reconstruction, have shown promise in medical imaging contexts \cite{shurrabSelfsupervisedLearningMethods2022,huangSelfsupervisedLearningMedical2023,chenSelfsupervisedLearningMedical2019}. Similarly, clustering-based approaches and unsupervised domain adaptation techniques attempt to segment images without explicit labels by leveraging inherent data structure or transferring knowledge from related domains \cite{tajbakhshSurrogateSupervisionMedical2019,peroneUnsupervisedDomainAdaptation2019,kumariDeepLearningUnsupervised2024}. However, these methods often struggle with the complex artefacts present in synchrotron data and may produce inconsistent results across time-series acquisitions where specimen morphology evolves significantly. Furthermore, the lack of ground truth validation makes it challenging to ensure segmentation accuracy for quantitative analyses.

Transfer learning approaches offer another promising direction for addressing the challenge of limited annotated data in synchrotron XCT segmentation. Transfer learning involves training a model on a source domain where data is abundant and transferring the learnt knowledge to a target domain where data is scarce \cite{kimTransferLearningMedical2022,karimiTransferLearningMedical2021}. Within this framework, domain adaptation techniques specifically address the distribution shift between source and target domains. These methods range from fine-tuning pre-trained models, to adversarial domain adaptation that learns domain-invariant features \cite{guanDomainAdaptationMedical2022,kanakasabapathyAdaptiveAdversarialNeural2021}, to source domain transformation where the source data is modified to better match the target domain characteristics \cite{tangGeneralizableFrameworkUnpaired2022}. In the context of XCT imaging, transfer learning could potentially leverage the abundance of high-quality laboratory XCT data to improve segmentation of artefact-laden synchrotron data. However, the substantial differences in image characteristics between these modalities---including variations in noise, contrast, and artefact types---present unique challenges for effective knowledge transfer.

Despite the advances in recent years, a significant gap remains in methodologies for efficiently training segmentation models for dynamic \textit{in situ} synchrotron XCT data without extensive manual annotation, particularly approaches that can effectively bridge the quality disparity between laboratory and synchrotron datasets while maintaining consistent performance throughout time-resolved experiments. In this work, we present a novel approach to overcome the challenge of limited training data for segmentation of synchrotron XCT data: leveraging more readily available \textit{ex situ} laboratory-based XCT data of similar specimens. Our methodology to achieve this involves preprocessing \textit{ex situ} XCT data to simulate key characteristics of synchrotron data, creating a robust training dataset for deep learning models. We demonstrate its effectiveness through the accurate segmentation of \textit{in situ} synchrotron XCT data of metal oxide specimens undergoing dissolution. The method achieves high accuracy on unseen data, and can be readily adapted for other materials systems and dynamic processes, potentially accelerating the analysis of time-resolved tomographic data across various scientific disciplines.

\section{Methodology}

\subsection{X-ray Computed Tomography Acquisition}

\paragraph{Copper Oxide Dissolution Study}
Two complementary X-ray computed tomography (XCT) approaches were employed in this study: high-quality laboratory-based scans for training data generation, and high-speed synchrotron imaging for capturing dissolution dynamics. 

Laboratory data were acquired using a Zeiss Xradia Versa 520 with parameters optimised for image quality: 140\,kV acceleration voltage, 10$\times$ objective magnification, and \SI{10}{\second} exposures. This configuration provided \SI{1.6}{\um} voxel size volumes (1024$\times$1024$\times$1024 voxels) with minimal artefacts and excellent contrast between the copper oxide phase and internal porosity.

Synchrotron imaging was performed at the I13-2 Diamond-Manchester Imaging beamline (Diamond Light Source, UK), where the high-brilliance monochromatic \SI{12}{\kilo\electronvolt} beam enabled rapid acquisition necessary for tracking dissolution kinetics. Using a pco.edge 5.5 detector (2560$\times$2160 pixels, \SI{6.5}{\um} pixel size) with 8$\times$ total magnification, we acquired 3000 projections over 180\textdegree{} with \SI{0.15}{\second} exposure time. This achieved \SI{0.8125}{\um} voxel size in reconstructed volumes of 2560$\times$2560$\times$2160 voxels, enabling acquisition of complete tomograms in approximately \SI{7.5}{\minute}. However, the rapid acquisition introduced several artefacts not present in the laboratory data, particularly ring artefacts and cupping effects.

The copper oxide specimens (approximately \SI{1}{\mm} diameter) underwent dissolution in 100~mM EDTA solution at pH~5 during continuous imaging, with approximately 200 tomograms acquired over 100 hours to capture the complete morphological evolution. Table~\ref{tab:xct_comparison} summarises the key differences between the two imaging modalities.

\begin{table}[h]
\centering
\caption{Comparison of laboratory and synchrotron XCT imaging parameters}
\label{tab:xct_comparison}
\begin{tabular}{lll}
\hline
\textbf{Parameter} & \textbf{Laboratory XCT} & \textbf{Synchrotron XCT} \\
\hline
Source & Polychromatic (140 kV) & Monochromatic (12 keV) \\
Detector & 1024$\times$1024 (2$\times$ binned) & 2560$\times$2160 \\
Voxel size & 1.6 \textmu m & 0.8125 \textmu m \\
Volume dimensions & 1024$\times$1024$\times$1024 & 2560$\times$2560$\times$2160 \\
Scan time & $\sim$5 hours & $\sim$7.5 minutes \\
Projections & 1400 over 360° & 3000 over 180° \\
Primary artefacts & Minimal & Ring artefacts, cupping \\
\hline
\end{tabular}
\end{table}

\subsection{Training Data Generation}
To train a neural network to segment poor quality synchrotron data, we developed a systematic ``data worsener" pipeline \cite{manchesterDataWorsener2025} to transform high-quality \textit{ex situ} data into realistic training examples that mimic synchrotron characteristics. The rationale for this approach, rather than training directly on limited synchrotron data, stems from two key advantages: (1) the abundance of available laboratory data (8,000 slices from eight specimens), and (2) the consistency of ground truth labels that can be automatically generated from artefact-free laboratory data using simple thresholding, avoiding the drift inherent in manual annotation over extended periods.

Each slice underwent a series of transformations designed to replicate synchrotron imaging artefacts:
\begin{itemize}
    \item \textbf{Initial processing}: Specimen segmentation via k-means clustering, followed by selective pore contrast reduction (pores below threshold reduced to 50\% intensity, blurred with a 5$\times$5 kernel)
    \item \textbf{Artefact creation}: Add edge cupping effects (100-pixel radius enhancement at 15,000 intensity units, with a 51$\times$51 blur kernel) and synthetic ring artefacts (80 rings with random intensities of 0.6--1.4$\times$ baseline and 1--2 pixel thickness).
    \item \textbf{Sinogram-space processing}: Radon transform (800 projections over 180°) followed by stripe removal using sorted median filtering (window size 200)\footnote{While seemingly counter-intuitive, this ring addition and removal process accurately reproduced the streaking artefacts observed in synchrotron data, as the removal algorithm deliberately leaves residual streaking patterns.}
    \item \textbf{Final reconstruction}: Filtered back projection (Shepp-Logan filter) and downsampling to 512\textsuperscript{3} volumes to enable reasonable batch sizes and training times given GPU memory constraints
\end{itemize}

The poor image quality observed in the synchrotron data arises not only from short exposure times resulting in higher signal-to-noise, but also from specimen movement during \textit{in situ} experiments and flow-induced vibrations as solution passes through the capillary setup. These factors compound the inherent artefacts from the imaging system itself. Parameters for the data worsener were experimentally optimised through iterative comparison with real synchrotron data until visual similarity was achieved.

This preprocessing strategy was applied to 8,000 slices from eight different specimens, creating a diverse training dataset that captured both the fundamental specimen features and the characteristic artefacts of synchrotron imaging. Each transformed volume was paired with its corresponding ground truth segmentation, derived from the original laboratory data using Otsu thresholding and validated through manual inspection of a random selection of slices. Otsu thresholding was suitable in this case as the laboratory data had high contrast and minimal artefacts.
\begin{figure}[h]
      \centering
      \begin{subfigure}[t]{0.32\textwidth}
        \centering
        \includegraphics[width=\textwidth]{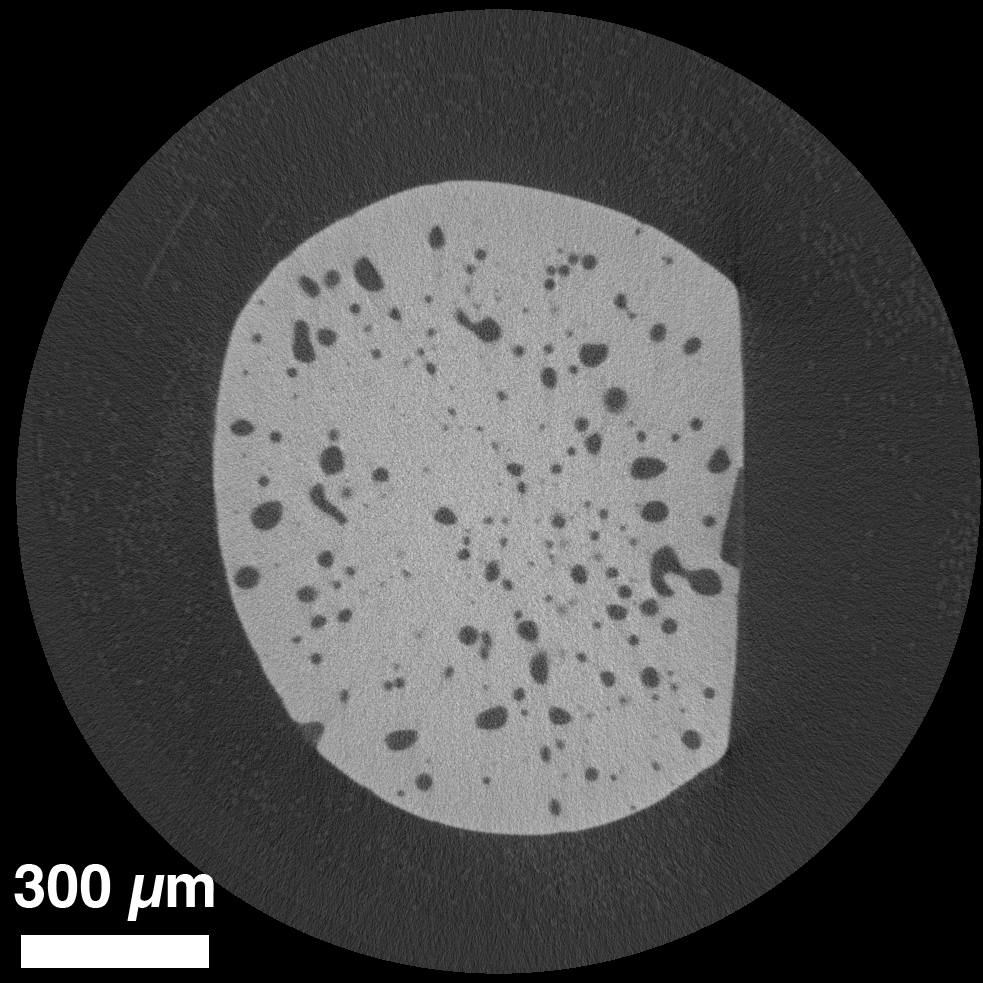}
      \end{subfigure}
      \hfill
      \begin{subfigure}[t]{0.32\textwidth}
        \centering
        \includegraphics[width=1\textwidth]{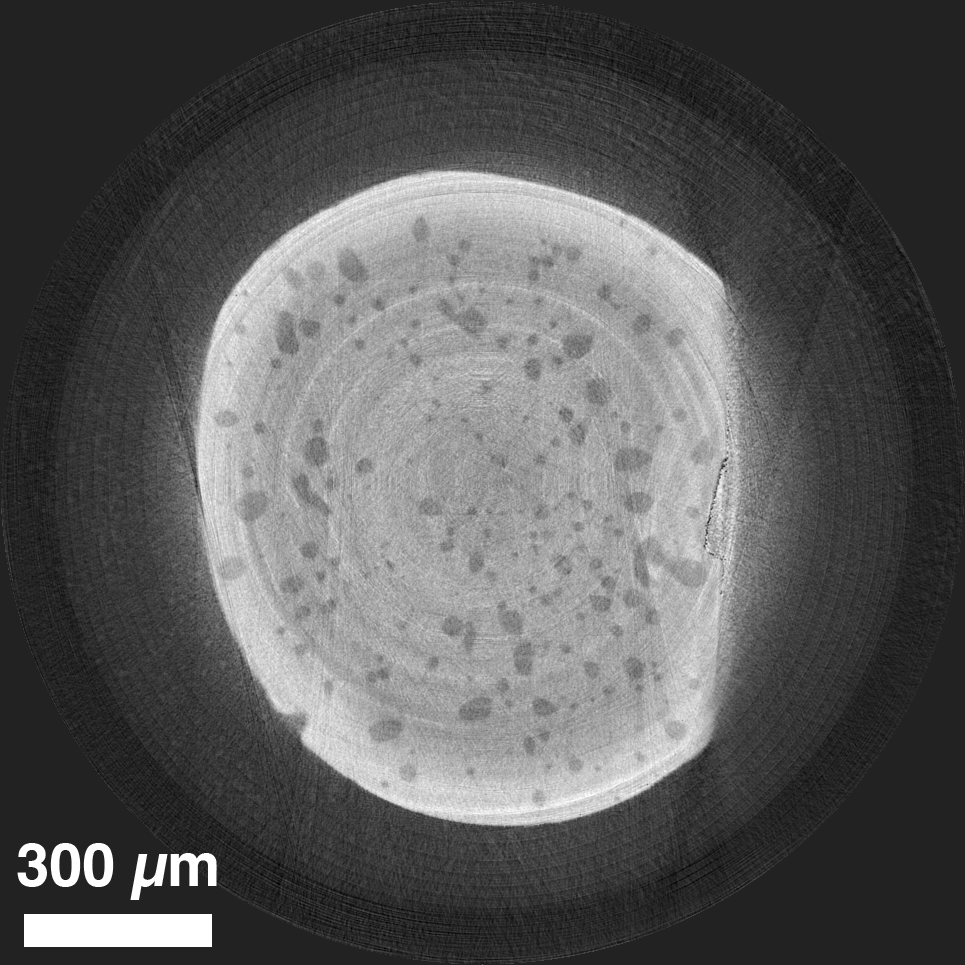}
        \label{fig:worsened_slice}
      \end{subfigure}
      \hfill
      \begin{subfigure}[t]{0.32\textwidth}
        \centering
        \includegraphics[width=1\textwidth]{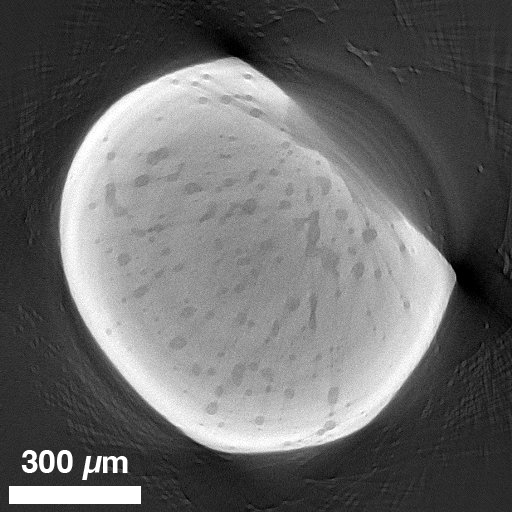}
      \end{subfigure}
\caption{Slice of lab-based \textit{ex situ} tomography data of a copper oxide pellet: (left) original slice, (middle) after post-processing with image processing techniques in Python to resemble \textit{in situ} data, and (right) synchrotron \textit{in situ} data for comparison.}
      \label{fig:slice_worsener}
\end{figure}
\subsection{Network Implementation}
Our segmentation pipeline was developed under significant computational constraints, using a laptop-based NVIDIA RTX 4070 (8~GB) GPU. These hardware limitations directly influenced our architectural choices and optimisation strategies, leading to a pragmatic balance between segmentation accuracy and computational efficiency.

The preprocessing pipeline extracts training data from XY, YZ, and XZ planes of each volume. This orientation-agnostic training strategy ensures the model learns to recognise features regardless of their orientation in the volume.

We employed the MiT-B4 variant of SegFormer \cite{xieSegFormerSimpleEfficient2021}, which comprises 60.8M parameters in the encoder and 3.3M parameters in the decoder. This configuration achieves an optimal balance between computational efficiency and segmentation accuracy, requiring 95.7 GFLOPs for inference while maintaining robust performance. The encoder utilises a hierarchical structure with progressively increasing channel dimensions [64, 128, 320, 512] across its four stages, enabling comprehensive multi-scale feature extraction.

When combined with the lightweight multi-layer perceptron (MLP) decoder head, the complete model contains 64.1M parameters total, substantially less than other state-of-the-art architectures such as SETR (318.3M) while achieving superior performance \cite{xieSegFormerSimpleEfficient2021}. Our implementation leverages the Hugging Face transformers library \cite{SegFormerHuggingFace}, using this backbone to generate features at multiple scales (1/4, 1/8, 1/16, and 1/32 of the original resolution).

To accommodate our single-channel tomographic data, we implemented a modification to the input layer of the pre-trained SegFormer model. The weights of the initial convolutional layer were adapted from three-channel RGB input to single-channel greyscale input by averaging the original RGB weights. This approach attempts to preserve the learnt feature detection capabilities of the pre-trained model:
\begin{lstlisting}[language=Python, label={lst:greyscale_conv}]
# Adapt first convolution layer for greyscale input
with torch.no_grad():
    # new_conv is the new single-channel conv
    # old_conv is the original pre-trained three-channel conv
    new_conv.weight.data = old_conv.weight.data.mean(dim=1, keepdim=True)
    if new_conv.bias is not None and old_conv.bias is not None:
        new_conv.bias.data = old_conv.bias.data
\end{lstlisting}

The SegFormer architecture \cite{xieSegFormerSimpleEfficient2021} inherently avoids traditional positional encodings. Instead, its hierarchical Mix Transformer (MiT) encoder utilises overlapped patch merging to generate multi-scale features. This design enhances local feature learning and allows the model to effectively process inputs of varying resolutions without the common issues of positional code interpolation. The overlapped patches inherently capture relative spatial information, which is beneficial for tasks like boundary detection in tomographic images. In our implementation, prior to fusing the multi-scale features in the decoder, we applied instance normalisation to each feature map. This step was taken to ensure consistent feature scales across different network depths:

\begin{lstlisting}[language=Python, label={lst:instance_norm}]
# Apply instance normalisation to encoder feature maps
hidden_states_normalized = [
    F.instance_norm(state)  # Normalise each channel independently
    for state in encoder_hidden_states  # Assuming hidden_states are from encoder
]
# These normalised states are then passed to the MLP decoder
\end{lstlisting}
SegFormer introduces an efficient self-attention mechanism with a sequence reduction process to manage computational overhead. The attention complexity is reduced from O(N\textsuperscript{2}) to O(N\textsuperscript{2}/R) where R is the reduction ratio applied to the token sequence length within each stage's self-attention computation. The reduction ratios are set to [64, 16, 4, 1] for the four hierarchical stages, meaning that within each stage, the token sequence length for self-attention is reduced by factors of 64, 16, 4, and 1 respectively from that stage's spatial resolution. This staged approach allows early layers to capture global context efficiently at lower computational cost, whilst later stages preserve fine-grained spatial details essential for accurate boundary delineation in semantic segmentation. This adaptation enables processing of large tomographic volumes while maintaining memory efficiency. The decoder employs a lightweight all-MLP design that unifies features from multiple scales through a series of linear projections and upsampling operations, comprising three main components:
\begin{itemize}
    \item Channel unification through MLP layers to standardise feature dimensions
    \item Hierarchical feature fusion with progressive upsampling
    \item Final MLP projection to generate binary or multi-label segmentation masks
\end{itemize}
The input volumes for the dissolution study were downsampled to 512$^3$ voxels and processed as unsigned 8-bit integers, reducing memory requirements while preserving essential morphological features and differentiation of pores with a similar grey value to their surroundings. While a single 512$^3$ volume at 8-bit precision requires only $\sim$134~MB, the 8~GB GPU memory constraint becomes limiting due to the model parameters ($\sim$256~MB), optimizer states ($\sim$512~MB for AdamW), gradients during backpropagation ($\sim$256~MB), and particularly the intermediate feature maps generated at multiple scales, which consume several GB for a batch size of 8. Binary cross-entropy dice loss (BCEDiceLoss) was employed as a loss function, suitable for the binary segmentation task at hand.

\subsection{Training Process}
Our training dataset of 8,000 slices from eight specimens was split at the slice level into training (80\%), validation (10\%), and test (10\%) sets. This slice-level splitting was necessary given the limited number of specimens available. Our training strategy incorporated comprehensive data augmentation techniques implemented through the Albumentations library \cite{buslaevAlbumentationsFastFlexible2020} to improve model robustness and generalisation. The augmentation pipeline was chosen to simulate common variations in synchrotron imaging data. We used brightness adjustments (range: $[-0.2, 0.2]$) and gamma corrections (range: $[90, 110]$) to mimic beam intensity fluctuations, while contrast adjustments and contrast limited adaptive histogram equalisation (CLAHE) addressed variations in image contrast. Geometric transformations, including random rotations and flips, further expanded the effective training dataset and ensured orientation invariance.

\begin{figure}[h]
      \centering
      \begin{subfigure}[t]{0.48\textwidth}
        \centering
        \includegraphics[width=\textwidth]{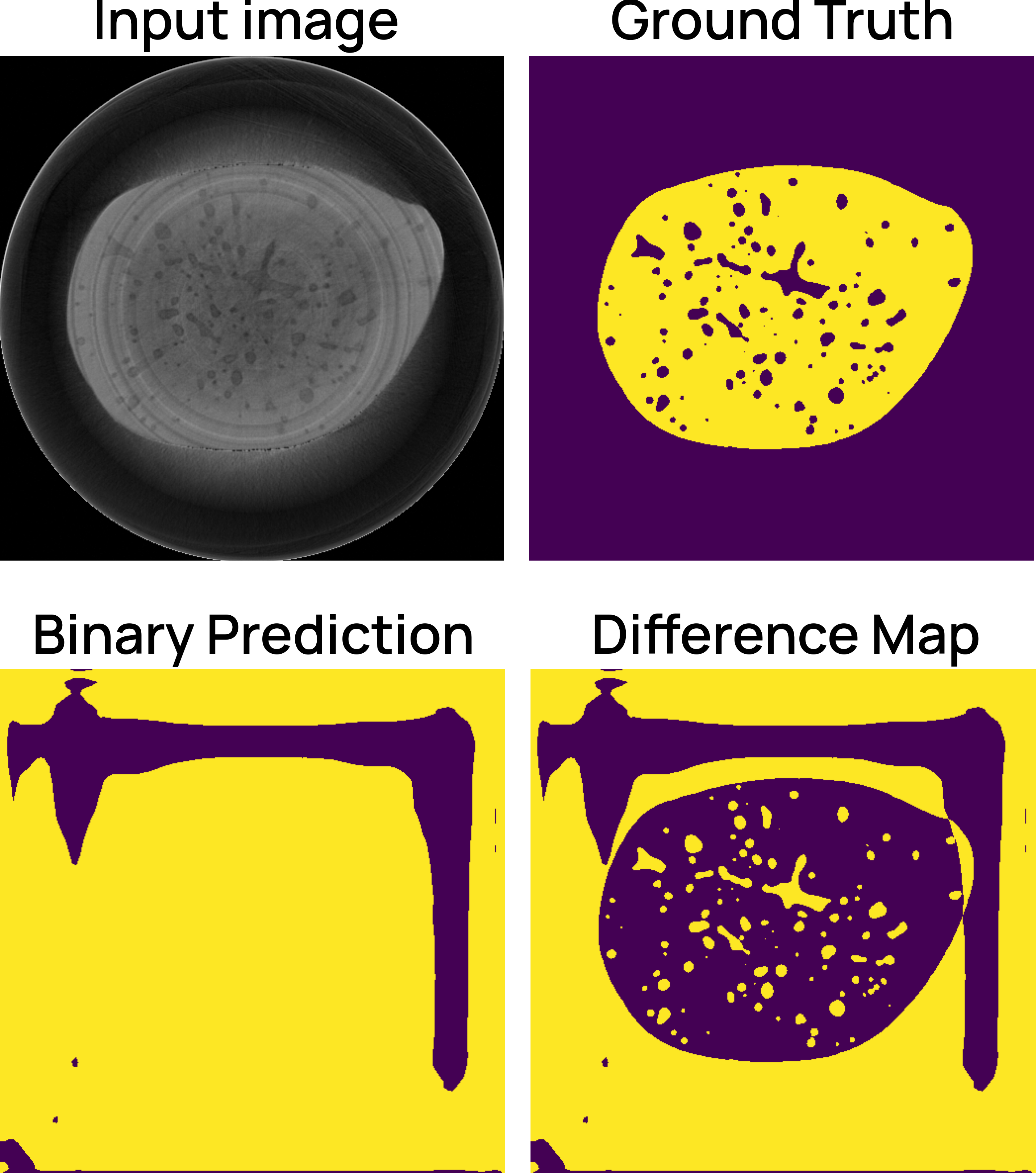}
        \caption{\centering Training visualisation for epoch 0}
      \end{subfigure}
      \hfill
      \vrule width 0.5pt
      \hfill
      \begin{subfigure}[t]{0.48\textwidth}
        \centering
        \includegraphics[width=1\textwidth]{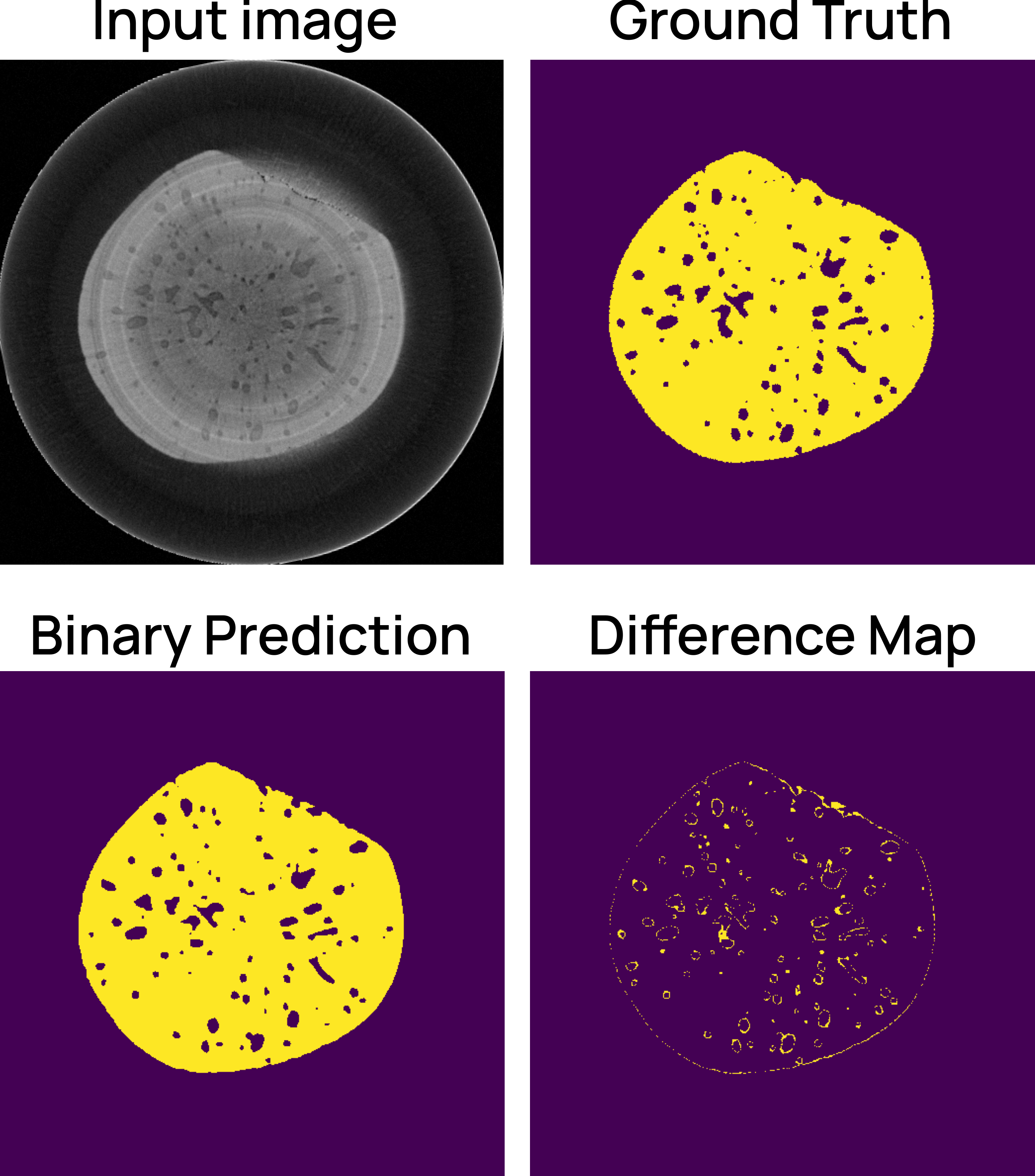}
        \caption{\centering Training visualisation for epoch 13}\label{fig:train_vis}
      \end{subfigure}
\caption{Visualisations produced during training at epoch 0 and epoch 13. Each panel displays four key components: (1) Input image: the preprocessed tomographic slice fed to the network; (2) Ground Truth (Target Segmentation): the reference binary mask derived from high-quality laboratory data; (3) Binary Prediction: the network's segmentation output after thresholding; and (4) Difference Map: highlighting discrepancies between the predicted segmentation and ground truth.}
      \label{fig:training_progress_examples}
\end{figure}

Training was carried out with a batch size of 8 slices---the maximum feasible given our GPU memory constraints of 8~GB. We used AdamW optimiser with an initial learning rate of 3.0e-7 and weight decay of 0.01, coupled with a OneCycleLR scheduler using a maximum learning rate of 2.5e-4 and a final dividing factor of 50. Training converged after 17 epochs, with each epoch requiring approximately 5 hours, reaching over 95\% mean intersection over union (mean IoU) on the test data (held-out worsened laboratory data) and 94.7\% mean IoU on previously unseen real synchrotron specimens vs. human manual segmentation. Figure~\ref{fig:train_vis} illustrates this rapid convergence, where visualisations from epoch 0 show significant segmentation errors (particularly visible in the difference map highlighting discrepancies between prediction and ground truth), while by epoch 13, the model produces accurate segmentations that closely match the target masks. The total duration of the training was approximately 3.5 days on a laptop grade NVIDIA RTX 4070 (which has significantly lower power and memory bandwidth than its desktop counterpart), demonstrating the feasibility of model retraining for adaptation to new experimental conditions or material systems. Despite these considerably constrained computational resources, our model exhibited robust generalisation across varying specimen morphologies and states, validating the effectiveness of our training approach.

\subsection{Human Annotation Reliability}
To establish performance benchmarks, we conducted inter-annotator and intra-annotator reliability analysis. Two expert researchers independently annotated a representative slice (slice 205 - close to the centre of the specimen), with one annotator re-segmenting after several days. Agreement was quantified using IoU and Cohen's kappa.

\section{Results}
\subsection{Segmentation Performance}
\begin{figure}[htbp]
    \centering
    \begin{subfigure}[t]{0.24\textwidth}
        \centering
        \includegraphics[width=\textwidth]{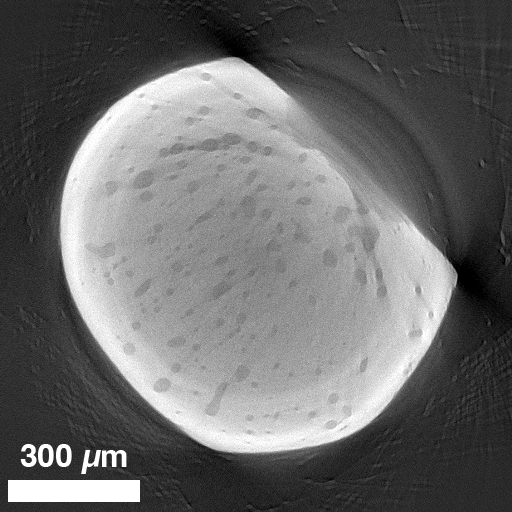}
        \caption{\centering Early stage reconstruction (xy)} 
    \end{subfigure}
    \hfill 
    \begin{subfigure}[t]{0.24\textwidth}
        \centering
        \includegraphics[width=\textwidth]{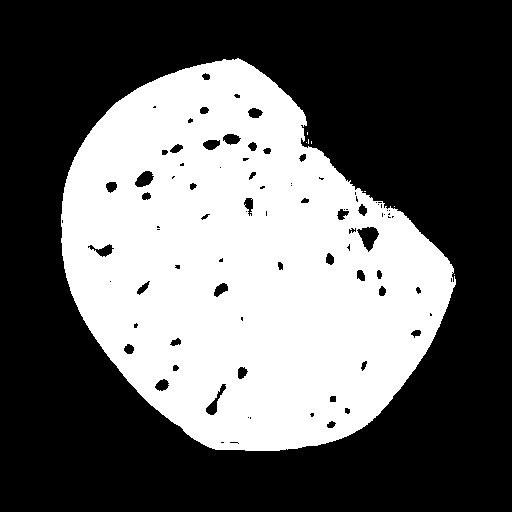}
        \caption{\centering Early stage seg. (xy)}
    \end{subfigure}
    \hfill 
    \begin{subfigure}[t]{0.24\textwidth}
        \centering
        \includegraphics[width=\textwidth]{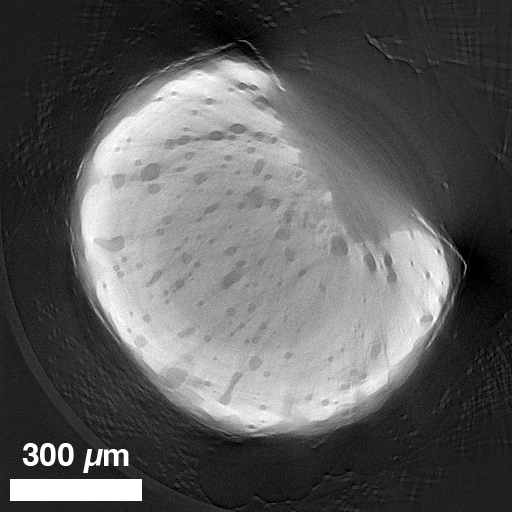}
        \caption{\centering Late stage reconstruction (xy)} 
    \end{subfigure}
    \hfill
    \begin{subfigure}[t]{0.24\textwidth}
        \centering
        \includegraphics[width=\textwidth]{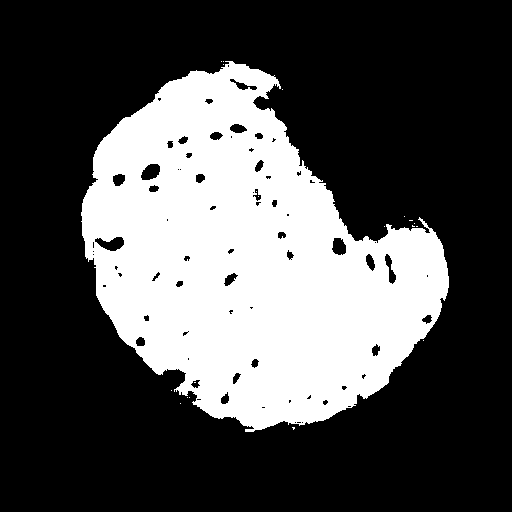}
        \caption{\centering Late stage seg. (xy)} 
    \end{subfigure}

    \vspace{\baselineskip}

    \begin{subfigure}[t]{0.24\textwidth}
        \centering
      
        \includegraphics[width=\textwidth]{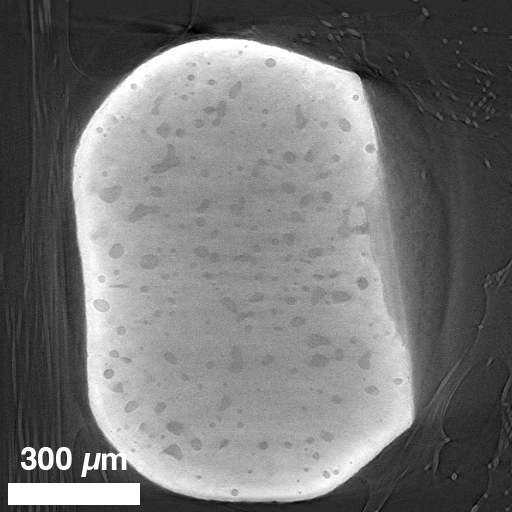}
        \caption{\centering Early stage reconstruction (yz)}
    \end{subfigure}
    \hfill
    \begin{subfigure}[t]{0.24\textwidth}
        \centering

        \includegraphics[width=\textwidth]{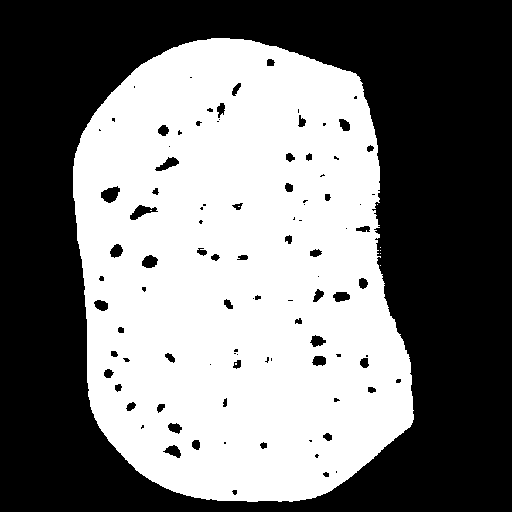}
        \caption{\centering Early stage seg. (yz)} 
    \end{subfigure}
    \hfill 
    \begin{subfigure}[t]{0.24\textwidth}
        \centering

        \includegraphics[width=\textwidth]{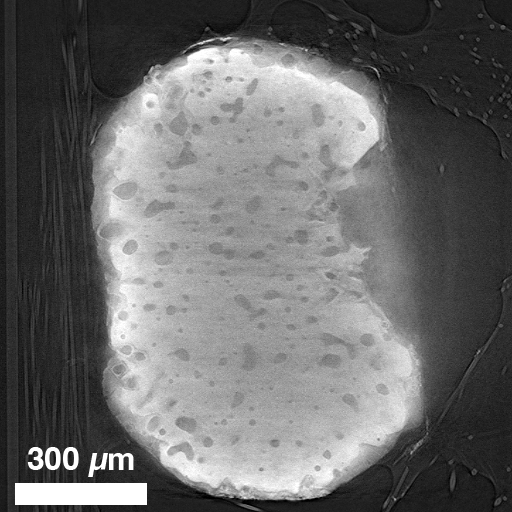}
        \caption{\centering Late stage reconstruction (yz)} 
    \end{subfigure}
    \hfill
    \begin{subfigure}[t]{0.24\textwidth}
        \centering
     
        \includegraphics[width=\textwidth]{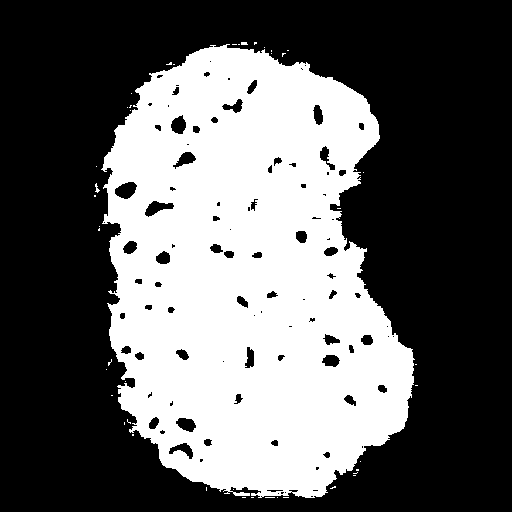}
        \caption{\centering Late stage seg. (yz)} 
    \end{subfigure}

    \caption{Neural network segmentation performance on synchrotron XCT copper oxide dissolution data in horizontal (xy, top row) and vertical (yz, bottom row) slices. (a, e) Early-stage reconstructed data. (b, f) Corresponding segmentations. (c, g) Late-stage reconstructed data showing morphological changes. (d, h) Corresponding segmentations, demonstrating consistent performance across evolution and orientations.}
    \label{fig:in_situ_prediction}
\end{figure}

Our neural network-based approach achieved robust segmentation of synchrotron tomography data, with mean IoU scores 94.7\% on unseen real synchrotron data versus human segmentation (Table \ref{tab:metrics}). The training converged after 17 epochs, showing consistent improvement in both training loss and validation metrics. This performance is particularly noteworthy given that the network was trained on high-quality laboratory XCT data yet successfully generalised to artefact-rich synchrotron data. Comparative examples with real data in Figure~\ref{fig:in_situ_prediction} demonstrate the model's ability to accurately delineate both internal porosity and specimen boundaries as the specimen's morphology evolves, though with some limitations in resolving pores below approximately 10 voxels ($\sim$23 \textmu m) in diameter, despite such features being present in the training data. 

Traditional segmentation methods such as Otsu thresholding are challenged by such data due to intensity variations and ring artefacts; in contrast, our approach maintained consistent performance across the entire dataset, processing 200 tomograms with an average segmentation time of approximately 30 seconds per volume. The model's most significant achievement was its ability to generalise across substantial morphological changes during dissolution. Despite training exclusively on intact specimens, the network successfully segmented specimens undergoing significant structural evolution. Performance remained consistent even when specimen morphology deviated significantly from the training examples, with mean IoU scores varying minimally between early and late dissolution stages (based on 6 hand-segmented slices in each case). This robustness suggests the network learned fundamental image features that remain stable throughout the dissolution process, rather than overfitting to specific morphological states.

\begin{table}[h]
\centering
\caption{Segmentation performance metrics across different evaluation sets}
\label{tab:metrics}
\begin{tabularx}{\textwidth}{lXXXXX} 
\hline
\textbf{Dataset} & \textbf{n} & \textbf{IoU (\%)} & \textbf{Precision (\%)} & \textbf{Recall (\%)} & \textbf{Dice (\%)} \\
\hline
\multicolumn{6}{l}{\textit{Transformed laboratory data}} \\
Validation & 1,165 & 97.0 ± 8.2 & 98.3 ± 5.2 & 98.6 ± 6.3 & 98.1 ± 7.6 \\
Test & 1,228 & 97.0 ± 7.4 & 98.1 ± 6.4 & 98.9 ± 3.5 & 98.3 ± 6.6 \\
\hline
\multicolumn{6}{l}{\textit{Real synchrotron data}} \\
Test (vs manual) & 12 & 94.7 ± 1.1 & 95.7 ± 1.1 & 98.9 ± 0.3 & 97.3 ± 0.6 \\
\hline
\multicolumn{6}{l}{\textit{Human benchmark}} \\
Inter-annotator & 1 & 94.6 & --- & --- & 97.2 \\
Intra-annotator & 1 & 96.5 & --- & --- & 98.2 \\
\hline
\end{tabularx}
\end{table}

\begin{figure}[htbp]
    \centering
    \begin{subfigure}[b]{0.48\textwidth}
        \centering
        \includegraphics[width=\textwidth]{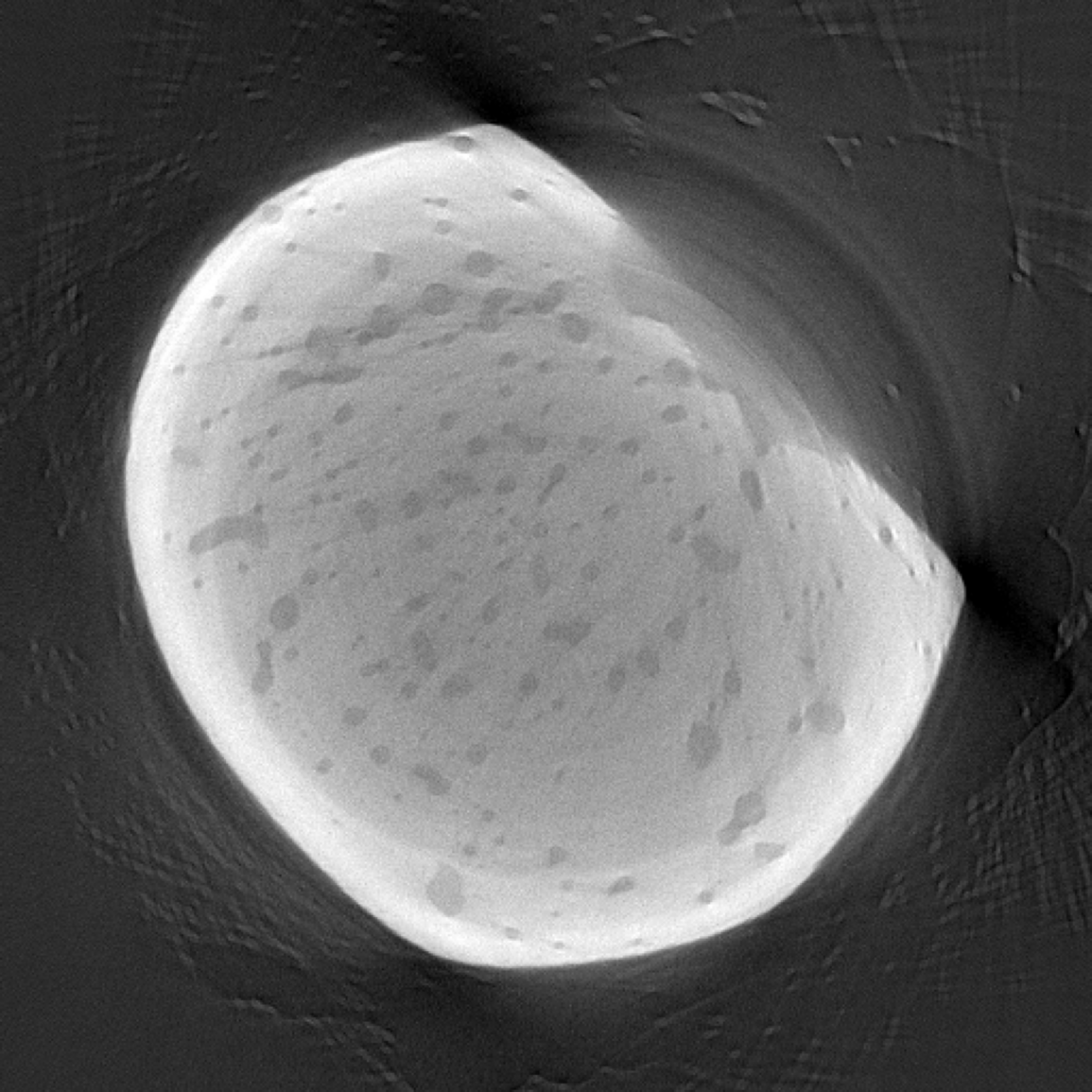}
        \caption{\centering Original data}
        \label{fig:seg_comparison_original}
    \end{subfigure}
    \hfill
    \begin{subfigure}[b]{0.48\textwidth}
        \centering
        \includegraphics[width=\textwidth]{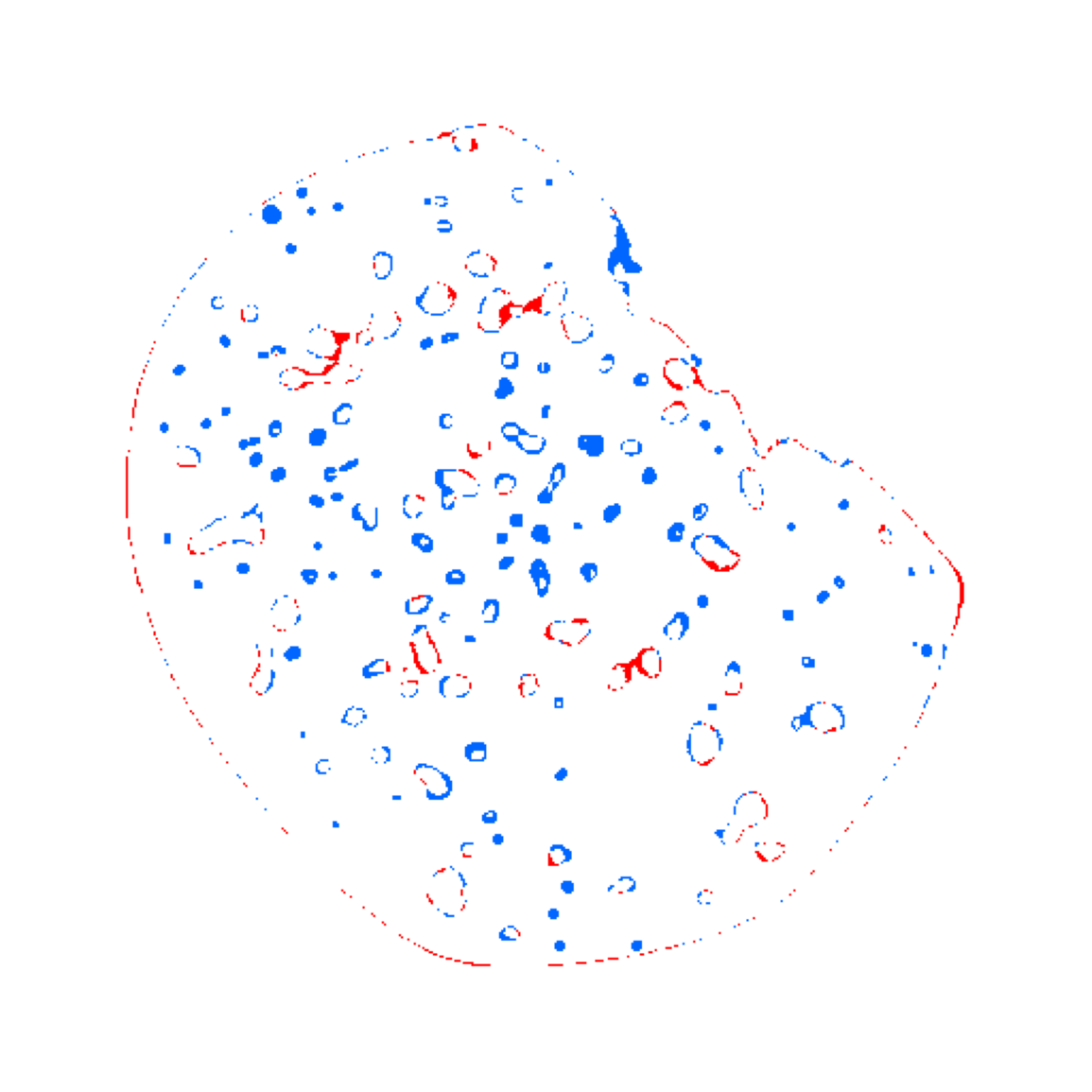}
        \caption{\centering Person 1 vs ML (IoU: 94.5\%)}
        \label{fig:seg_comparison_person1_ml}
    \end{subfigure}
    
    \vspace{\baselineskip}
    
    \begin{subfigure}[b]{0.48\textwidth}
        \centering
        \includegraphics[width=\textwidth]{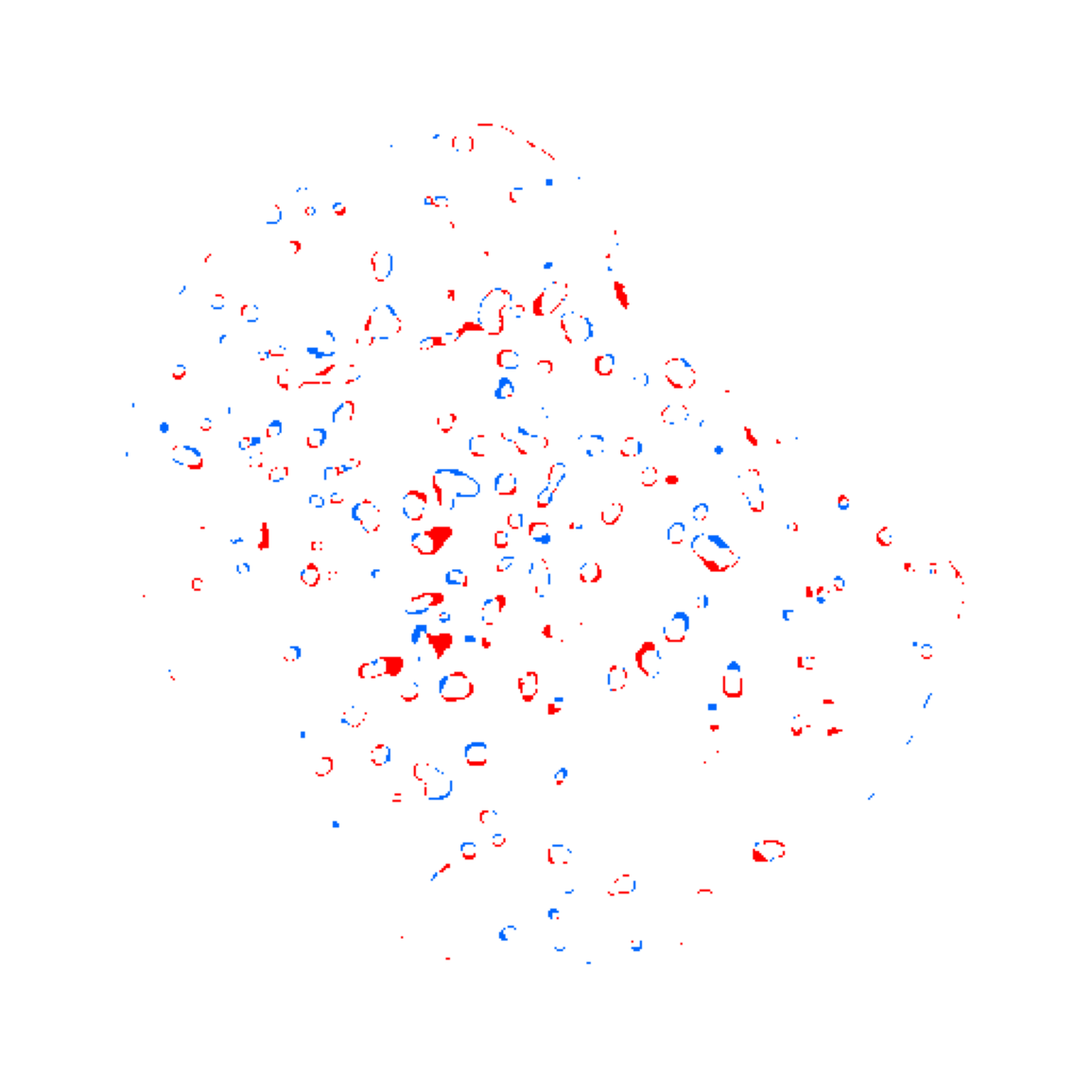}
        \caption{\centering Person 1 vs Person 1 (attempt 2) (IoU: 96.5\%)}
        \label{fig:seg_comparison_person1_attempt2}
    \end{subfigure}
    \hfill
    \begin{subfigure}[b]{0.48\textwidth}
        \centering
        \includegraphics[width=\textwidth]{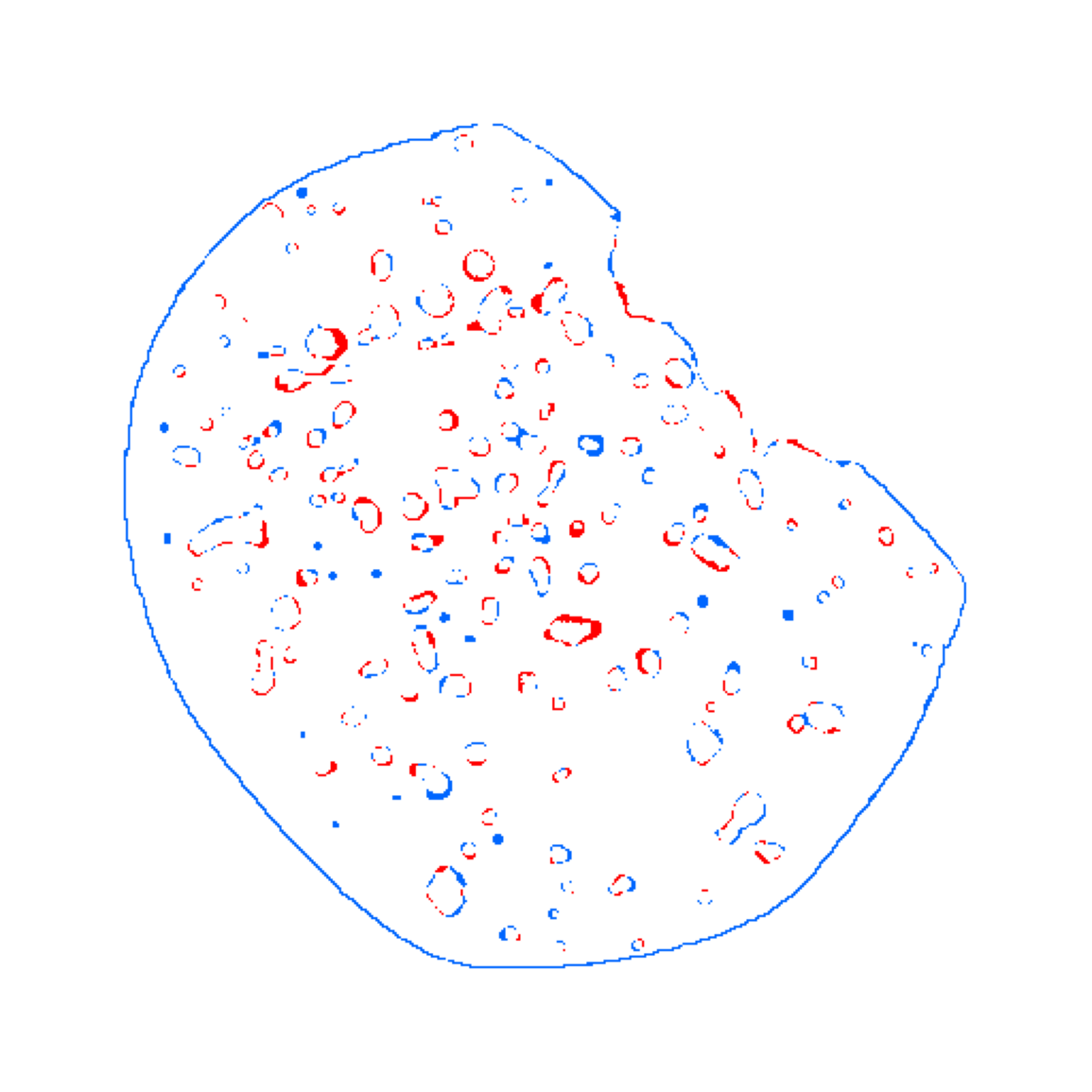}
        \caption{\centering Person 1 vs Person 2 (IoU: 94.6\%)}
        \label{fig:seg_comparison_person1_person2}
    \end{subfigure}
    
    \caption{Segmentation comparison showing inter-annotator agreement for a single slice of synchrotron data. (a) Original reconstructed data. (b-d) Difference maps where red pixels indicate regions marked by the first annotator but not the second, blue pixels indicate regions marked by the second annotator but not the first, and white indicates agreement. The high IoU scores demonstrate strong agreement between human annotators and between humans and the ML model, with slightly higher consistency within the same human annotator (0.965) compared to inter-human (0.946) or human-AI (0.945) comparisons. Notable is the difference in edge-finding tactic used between human annotators, with a consistent 1--2 pixel difference at the outer edge of the specimen.}
    \label{fig:segmentation_comparison}
\end{figure}

To contextualise our AI performance, we quantified human annotation reliability. Agreement between two expert annotators yielded 94.6\% IoU (Cohen's $\kappa$ = 0.957), whilst intra-annotator reliability achieved 96.5\% IoU ($\kappa$ = 0.972). Notably, our AI model's performance (94.7\% IoU) matched human inter-annotator agreement, with AI-human agreement showing excellent reliability ($\kappa$ = 0.951$\pm$0.005). Figure \ref{fig:segmentation_comparison} shows inter and intra-annotator segmentation versus the ML segmentation. These results indicate our approach has reached the practical performance ceiling for this segmentation task.

\section{Discussion}
The \textit{ex situ} transformation approach demonstrated in this study represents a significant innovation in automating the segmentation of \textit{in situ} synchrotron tomography data. Using readily available laboratory XCT datasets and systematically transforming them to mimic synchrotron imaging characteristics, we effectively bridged the quality gap between these two data sources without requiring extensive manual segmentation.

The performance of our SegFormer-based model demonstrates that deep learning approaches can successfully generalise from transformed laboratory data to authentic synchrotron data, even when significant imaging artefacts are present. This generalisation capacity is particularly evident in the model's consistent performance across different specimens and morphological states during dissolution. The high mean IoU scores achieved on unseen data validate the effectiveness of our training data generation approach.\\
A key finding is the model's ability to maintain segmentation accuracy throughout the dissolution process, despite being trained exclusively on intact specimens. This suggests that the network has learnt fundamental image features and specimen characteristics that remain identifiable even as the copper oxide undergoes significant structural evolution. Such robustness to morphological changes is crucial for time-series experiments, where consistent segmentation across all time points is essential for reliable quantitative analysis.\\
The computational efficiency of our approach represents another significant advantage. While traditional segmentation methods might require several hours to process a single tomographic volume, our model completes the same task in approximately 30 seconds. This 100-fold reduction in processing time transforms what was previously a substantial bottleneck in the experimental workflow, enabling near real-time analysis during beamtime experiments. Furthermore, the consistent segmentation quality eliminates the variability inherent in manual approaches, improving the reproducibility of quantitative analyses.
\begin{figure}[h!] 
    \centering
    \begin{subfigure}[t]{0.48\textwidth} 
        \centering
        \includegraphics[width=\textwidth]{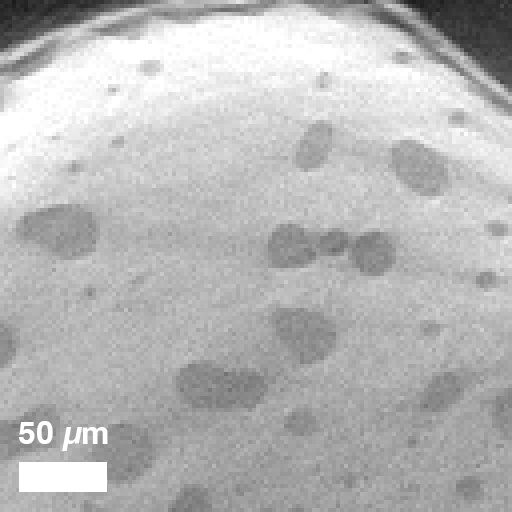}
        \caption{\centering Reconstructed data region}
    \end{subfigure}
    \hspace{0.02\textwidth}
    \begin{subfigure}[t]{0.48\textwidth} 
        \centering
        \includegraphics[width=\textwidth]{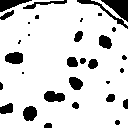}
        \caption{\centering Manual segmentation (Ground Truth)}
    \end{subfigure}

    \vspace{\baselineskip} 

    \begin{subfigure}[t]{0.48\textwidth}
        \centering
        \includegraphics[width=\textwidth]{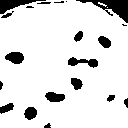}
        \caption{\centering Network prediction}
    \end{subfigure}
    \hspace{0.02\textwidth}
    \begin{subfigure}[t]{0.48\textwidth} 
        \centering
        \includegraphics[width=\textwidth]{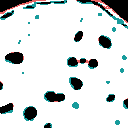}
        \caption{\centering Difference map (Network vs Manual)}
    \end{subfigure}
    \caption{Detailed comparison of segmentation performance in a representative zoomed-in region. (a) Reconstructed synchrotron XCT data. (b) Manual segmentation reference. (c) Segmentation predicted by the neural network. (d) Difference map highlighting discrepancies: Blue regions indicate areas incorrectly labelled 1, e.g., pore edges/small pores; Red regions indicate areas incorrectly labelled as 0, e.g., outer edge, filled gaps.}
    \label{fig:detailed_comparison}
\end{figure}
A closer look at segmentation quality against manual annotations reveals subtle patterns (Figure~\ref{fig:detailed_comparison}). The network tends to slightly undersegment pore edges (blue regions in d), making them appear smaller, and struggles with pores below ~10 voxels ($\sim$23 \textmu m). It also occasionally oversegments the specimen's outer boundary or fills narrow gaps (red regions). These systematic boundary errors account for the 5.3\% IoU discrepancy, rather than major failures. Understanding the specific error modes helps us interpret quantitative results derived from the segmentation and informs potential avenues for future model refinement.\\
The finding that neural network performance (94.7\% IoU, n=12 slices) closely matches human inter-annotator reliability (94.6\% IoU, n=1 slice) provides important context for interpreting our results. While the limited sample size for human reliability assessment warrants further validation, this preliminary comparison suggests the 3.4 percentage point difference between validation (98.1\%) and test performance (94.7\%) may reflect inherent annotation uncertainty in complex synchrotron data rather than model limitations. Furthermore, intra-annotator metrics for the same slice (IoU 96.5\%) illustrate a limitation of human semantic segmentation: results are not deterministic and may vary significantly day to day, whereas a trained model will always produce the same results for a particular labelling task. 

\subsection{Limitations and Future Work}
While our approach performed well, several limitations suggest avenues for improvement.\\
Firstly, it required reasonably large computational resources, particularly for network training. Our implementation used a consumer-grade GPU (NVIDIA RTX 4070, 8 GB), requiring approximately 3.5 days of training time. For researchers without access to dedicated GPU hardware, this represents a significant barrier to adoption. However, this computational cost is highly hardware-dependent; training on a high-end consumer GPU could reduce this time to a few hours, while a data-center GPU (e.g., NVIDIA H100) could complete the task in under an hour, primarily by enabling much larger batch sizes.\\
The slice-by-slice segmentation strategy, while computationally efficient, discards valuable 3D contextual information that could improve segmentation coherence across the volume. Although we implemented a 12-axis prediction approach (3 orthogonal planes, 4 rotations each) with ensemble voting to mitigate this limitation, methods like 3D convolutions or "2.5D" approaches (where neighbouring slices are used for a prediction) might better preserve volumetric continuity. Our approach also demonstrated reduced accuracy when segmenting features smaller than approximately 10 voxels in diameter ($\sim$23 \textmu m), particularly in regions with severe imaging artefacts. Furthermore, minor systematic inaccuracies were observed at boundaries, including slight undersegmentation of internal pore edges and oversegmentation of the specimen's outer edge (Figure~\ref{fig:detailed_comparison}).\\
The \textit{ex situ} transformation approach crucially depends on access to laboratory-based XCT facilities capable of producing high-quality reference scans. This reliance creates both logistical and financial barriers, as laboratory scanning services often have limited availability and significant associated costs. Furthermore, this methodology assumes sufficient similarity between \textit{ex situ} and \textit{in situ} specimen morphology at the initial state to establish valid training examples. For materials that undergo significant changes between laboratory and synchrotron environments, this assumption may not hold.\\
Several promising directions could address these limitations and extend the capabilities of deep learning for tomographic segmentation. For time-resolved studies, incorporating temporal consistency constraints could further improve segmentation accuracy by leveraging the gradual nature of most morphological changes. Transfer learning strategies also merit investigation, potentially allowing models trained on one material system to be efficiently fine-tuned for another.\\
Furthermore, exploring alternative loss functions could potentially mitigate some of the observed segmentation inaccuracies. For instance, employing losses that directly optimise the Intersection over Union (IoU) metric, such as the Lovász-Softmax loss \cite{bermanLovaszSoftmaxLossTractable2018}, or using functions designed to handle class imbalance and boundary definition like Focal loss \cite{linFocalLossDense2020} or Tversky loss \cite{salehiTverskyLossFunction2017}, might improve performance on small features and enhance boundary delineation compared to the BCEDiceLoss used in this study.\\
This approach could significantly reduce the training data requirements for new applications, making the methodology more accessible to researchers with limited computational resources. Active learning frameworks, where the model identifies areas of high uncertainty for targeted manual annotation, could optimise the human effort required for developing training datasets.\\
Furthermore, while this study demonstrated binary segmentation, the underlying network architecture inherently supports multi-phase tasks. Future work could therefore explore applying this methodology, or adaptations thereof, to segment more complex materials systems, which would likely require developing specific training data strategies suitable for multi-component analysis.

\section{Conclusion}
In this work, we presented a methodology for addressing the challenge of automated segmentation in synchrotron X-ray computed tomography. The approach of transforming high-quality \textit{ex situ} laboratory data to resemble poorer quality \textit{in situ} data of the same samples successfully overcame the fundamental paradox of needing extensive labelled training data to avoid manual segmentation. Our implementation demonstrated good accuracy across a distinct material system and experimental scenario: \textit{in situ} copper oxide dissolution under aqueous conditions.\\
We used readily available laboratory data, applying systematic modifications to simulate synchrotron imaging characteristics while preserving ground-truth segmentation information. This tactic achieved high segmentation accuracy on the target datasets, while reducing processing time by 1--2 orders of magnitude.\\
Perhaps most significantly, the approach demonstrated exceptional generalisability across significant morphological changes. The copper oxide model successfully segmented specimens that underwent dissolution-induced evolution, despite training exclusively on intact specimens.\\
This methodology offers practical solutions for the analysis of time-resolved tomographic data in a variety of materials systems. The open source implementation provides a flexible framework that can be readily adapted to diverse material systems and experimental conditions, potentially accelerating scientific discovery by removing the analysis bottleneck often associated with large-scale tomographic studies.\\
As synchrotron facilities continue to advance detector technology and acquisition speeds, the resulting increase in data volume will only intensify the need for automated analysis pipelines. The approaches presented here represent important steps towards addressing this challenge, enabling researchers to fully leverage the temporal resolution afforded by modern synchrotron beamlines to investigate dynamic processes across multiple length and time scales.

\section*{Acknowledgements}
This work was carried out with the support of Diamond Light Source, instrument I13-2 (proposal MG31912). The authors thank Dr Slava Kachkanov (Diamond Light Source) for his assistance and support during the experiment conducted at the I13-2 beamline, and Dr Elizabeth Parker-Quaife, Dr Robert Burrows, and the United Kingdom National Nuclear Laboratory for their support and supervision during this research.

This research was part of a doctoral project funded by Engineering and Physical Sciences Research Council (EPSRC) grant EP/L01680X/1, and United Kingdom National Nuclear Laboratory (UKNNL) grant NNL/UA/094.

The authors acknowledge the use of Anthropic Claude 3.7 (Anthropic, via claude.com) for assistance with refining the writing and correcting grammar during the preparation of this manuscript. All AI-generated suggestions were reviewed and edited by the authors, who take full responsibility for the final content of this publication.

\section*{Code Availability}
The code for this project is available via the author's Github page \cite{manchesterDataWorsener2025,manchesterScrambledSeg2025}.

\bibliography{refs_better_bibtex}

\begin{thebibliography}{10}

\bibitem{withersXrayComputedTomography2021}
Philip~J. Withers, Charles Bouman, Simone Carmignato, Veerle Cnudde, David Grimaldi, Charlotte~K. Hagen, Eric Maire, Marena Manley, Anton Du~Plessis, and Stuart~R. Stock.
\newblock X-ray computed tomography.
\newblock {\em Nature Reviews Methods Primers}, 1(1):1--21, February 2021.

\bibitem{fineganInoperandoHighspeedTomography2015}
Donal~P. Finegan, Mario Scheel, James~B. Robinson, Bernhard Tjaden, Ian Hunt, Thomas~J. Mason, Jason Millichamp, Marco Di~Michiel, Gregory~J. Offer, Gareth Hinds, Dan~J.L. Brett, and Paul~R. Shearing.
\newblock In-operando high-speed tomography of lithium-ion batteries during thermal runaway.
\newblock {\em Nature Communications}, 6(1):6924, April 2015.

\bibitem{loweMetamorphosisRevealedTimelapse2013}
Tristan Lowe, Russell~J. Garwood, Thomas~J. Simonsen, Robert~S. Bradley, and Philip~J. Withers.
\newblock Metamorphosis revealed: Time-lapse three-dimensional imaging inside a living chrysalis.
\newblock {\em Journal of The Royal Society Interface}, 10(84):20130304, July 2013.

\bibitem{vanoffenwertPoreScaleVisualizationQuantification2019}
Stefanie Van~Offenwert, Veerle Cnudde, and Tom Bultreys.
\newblock Pore-{{Scale Visualization}} and {{Quantification}} of {{Transient Solute Transport Using Fast Microcomputed Tomography}}.
\newblock {\em Water Resources Research}, 55(11):9279--9291, November 2019.

\bibitem{chao-kungyangSimulationStudyMotion1982}
{Chao-Kung Yang}, {S C Orphanoudakis}, and {J W Strohbehn}.
\newblock A simulation study of motion artefacts in computed tomography.
\newblock {\em Physics in Medicine \& Biology}, 27(1):51--61, January 1982.

\bibitem{toftsSourcesArtefactComputed1980}
P~S Tofts and J~C Gore.
\newblock Some sources of artefact in computed tomography.
\newblock {\em Physics in Medicine \& Biology}, 25(1):117--127, January 1980.

\bibitem{ohnesorgeEfficientCorrectionCT2000}
B.~Ohnesorge, T.~Flohr, K.~Schwarz, J.~P. Heiken, and K.~T. Bae.
\newblock Efficient correction for {{CT}} image artifacts caused by objects extending outside the scan field of view.
\newblock {\em Medical Physics}, 27(1):39--46, January 2000.

\bibitem{rabrooksBeamHardeningXray1976}
{R A Brooks} and {G Di Chiro}.
\newblock Beam hardening in {{X-ray}} reconstructive tomography.
\newblock {\em Physics in Medicine \& Biology}, 21(3):390--398, May 1976.

\bibitem{kakPrinciplesComputerizedTomographic2001}
Avinash~C. Kak and Malcolm Slaney.
\newblock {\em Principles of Computerized Tomographic Imaging: "{{This SIAM}} Edition Is an Unabridged Republication of the Work First Published by {{IEEE Press}}, {{New York}}, 1988."}.
\newblock Number~33 in Classics in Applied Mathematics. {Society for Industrial and Applied Mathematics (SIAM, 3600 Market Street, Floor 6, Philadelphia, PA 19104)}, Philadelphia, Pa, 2001.

\bibitem{sheppardTechniquesImageEnhancement2004}
Adrian~P. Sheppard, Robert~M. Sok, and Holger Averdunk.
\newblock Techniques for image enhancement and segmentation of tomographic images of porous materials.
\newblock {\em Physica A: Statistical Mechanics and its Applications}, 339(1-2):145--151, August 2004.

\bibitem{torralbaComparisonSurfaceExtraction2018}
Marta Torralba, Roberto Jim{\'e}nez, Jos{\'e}~A. {Yag{\"u}e-Fabra}, Sinu{\'e} Ontiveros, and Guido Tosello.
\newblock Comparison of surface extraction techniques performance in computed tomography for {{3D}} complex micro-geometry dimensional measurements.
\newblock {\em The International Journal of Advanced Manufacturing Technology}, 97(1-4):441--453, July 2018.

\bibitem{borgesdeoliveiraExperimentalInvestigationSurface2016}
Fabr{\'i}cio Borges De~Oliveira, Alessandro Stolfi, Markus Bartscher, Leonardo De~Chiffre, and Ulrich {Neuschaefer-Rube}.
\newblock Experimental investigation of surface determination process on multi-material components for dimensional computed tomography.
\newblock {\em Case Studies in Nondestructive Testing and Evaluation}, 6:93--103, November 2016.

\bibitem{sokacImprovedSurfaceExtraction2020}
Mario Sokac, Igor Budak, Marko Katic, Zivana Jakovljevic, Zeljko Santosi, and Djordje Vukelic.
\newblock Improved surface extraction of multi-material components for single-source industrial {{X-ray}} computed tomography.
\newblock {\em Measurement}, 153:107438, March 2020.

\bibitem{ronnebergerUNetConvolutionalNetworks2015}
Olaf Ronneberger, Philipp Fischer, and Thomas Brox.
\newblock U-{{Net}}: {{Convolutional Networks}} for {{Biomedical Image Segmentation}}.
\newblock In Nassir Navab, Joachim Hornegger, William~M. Wells, and Alejandro~F. Frangi, editors, {\em Medical {{Image Computing}} and {{Computer-Assisted Intervention}} -- {{MICCAI}} 2015}, Lecture {{Notes}} in {{Computer Science}}, pages 234--241, Cham, 2015. Springer International Publishing.

\bibitem{strohmannSemanticSegmentationSynchrotron2019}
Tobias Strohmann, Katrin Bugelnig, Eric Breitbarth, Fabian Wilde, Thomas Steffens, Holger Germann, and Guillermo Requena.
\newblock Semantic segmentation of synchrotron tomography of multiphase {{Al-Si}} alloys using a convolutional neural network with a pixel-wise weighted loss function.
\newblock {\em Scientific Reports}, 9(1), December 2019.

\bibitem{silveiraDeepLearningOvercome2024}
Andreia Silveira, Imke Greving, Elena Longo, Mario Scheel, Timm Weitkamp, Claudia Fleck, Ron Shahar, and Paul Zaslansky.
\newblock Deep learning to overcome {{Zernike}} phase-contrast {{nanoCT}} artifacts for automated micro-nano porosity segmentation in bone.
\newblock {\em Journal of Synchrotron Radiation}, 31(1):136--149, January 2024.

\bibitem{liAMSegNetAdditiveManufacturing2024}
Wei Li, Rub{\'e}n {Lambert-Garcia}, Anna C.~M. Getley, Kwan Kim, Shishira Bhagavath, Marta Majkut, Alexander Rack, Peter~D. Lee, and Chu Lun~Alex Leung.
\newblock {{AM-SegNet}} for additive manufacturing {\emph{in situ}} {{X-ray}} image segmentation and feature quantification.
\newblock {\em Virtual and Physical Prototyping}, 19(1), December 2024.

\bibitem{shiMultistageDeepLearning2025}
Jiayang Shi, Dani{\"e}l~M. Pelt, and K.~Joost Batenburg.
\newblock Multi-stage deep learning artifact reduction for parallel-beam computed tomography.
\newblock {\em Journal of Synchrotron Radiation}, 32(2):442--456, March 2025.

\bibitem{haoDeployingMachineLearning2023}
Guanhua Hao, Eric~J. Roberts, Tanny Chavez, Zhuowen Zhao, Elizabeth~A. Holman, Howard Yanxon, Adam Green, Harinarayan Krishnan, Daniela Ushizima, Dylan McReynolds, Nicholas Schwarz, Petrus~H. Zwart, Alexander Hexemer, and Dilworth~Y. Parkinson.
\newblock Deploying {{Machine Learning Based Segmentation}} for {{Scientific Imaging Analysis}} at {{Synchrotron Facilities}}.
\newblock {\em IS\&T International Symposium on Electronic Imaging}, 35:IPAS--290, 2023.

\bibitem{usmanakbarBrainTumorSegmentation2024}
Muhammad Usman~Akbar, M{\aa}ns Larsson, Ida Blystad, and Anders Eklund.
\newblock Brain tumor segmentation using synthetic {{MR}} images - {{A}} comparison of {{GANs}} and diffusion models.
\newblock {\em Scientific Data}, 11(1):259, February 2024.

\bibitem{thambawitaSinGANSegSyntheticTraining2022}
Vajira Thambawita, Pegah Salehi, Sajad~Amouei Sheshkal, Steven~A. Hicks, Hugo~L. Hammer, Sravanthi Parasa, Thomas~De Lange, P{\aa}l Halvorsen, and Michael~A. Riegler.
\newblock {{SinGAN-Seg}}: {{Synthetic}} training data generation for medical image segmentation.
\newblock {\em PLOS ONE}, 17(5):e0267976, May 2022.

\bibitem{koetzierGeneratingSyntheticData2024}
Lennart~R. Koetzier, Jie Wu, Domenico Mastrodicasa, Aline Lutz, Matthew Chung, W.~Adam Koszek, Jayanth Pratap, Akshay~S. Chaudhari, Pranav Rajpurkar, Matthew~P. Lungren, and Martin~J. Willemink.
\newblock Generating {{Synthetic Data}} for {{Medical Imaging}}.
\newblock {\em Radiology}, 312(3):e232471, September 2024.

\bibitem{fengEnhancingMedicalImaging2024}
Yinqiu Feng, Bo~Zhang, Lingxi Xiao, Yutian Yang, Tana Gegen, and Zexi Chen.
\newblock Enhancing {{Medical Imaging}} with {{GANs Synthesizing Realistic Images}} from {{Limited Data}}.
\newblock In {\em 2024 {{IEEE}} 4th {{International Conference}} on {{Electronic Technology}}, {{Communication}} and {{Information}} ({{ICETCI}})}, pages 1192--1197, Changchun, China, May 2024. IEEE.

\bibitem{sardharaGenerativeAdversarialNetworks2025}
Trushal Sardhara, Christian Johannes~J Cyron, Martin Ritter, and Roland~C Aydin.
\newblock Generative adversarial networks for creating realistic training data for machine learning-based segmentation of {{FIB}} tomography data.
\newblock {\em Machine Learning: Science and Technology}, April 2025.

\bibitem{huDiscriminatorCooperatedFeatureMap2023}
Tie Hu, Mingbao Lin, Lizhou You, Fei Chao, and Rongrong Ji.
\newblock Discriminator-{{Cooperated Feature Map Distillation}} for {{GAN Compression}}.
\newblock In {\em 2023 {{IEEE}}/{{CVF Conference}} on {{Computer Vision}} and {{Pattern Recognition}} ({{CVPR}})}, pages 20351--20360, Vancouver, BC, Canada, June 2023. IEEE.

\bibitem{wangQGANQuantizedGenerative2019}
Peiqi Wang, Dongsheng Wang, Yu~Ji, Xinfeng Xie, Haoxuan Song, XuXin Liu, Yongqiang Lyu, and Yuan Xie.
\newblock {{QGAN}}: {{Quantized Generative Adversarial Networks}}, 2019.

\bibitem{saxenaGenerativeAdversarialNetworks2022}
Divya Saxena and Jiannong Cao.
\newblock Generative {{Adversarial Networks}} ({{GANs}}): {{Challenges}}, {{Solutions}}, and {{Future Directions}}.
\newblock {\em ACM Computing Surveys}, 54(3):1--42, April 2022.

\bibitem{chenChallengesCorrespondingSolutions2021}
Haiyang Chen.
\newblock Challenges and {{Corresponding Solutions}} of {{Generative Adversarial Networks}} ({{GANs}}): {{A Survey Study}}.
\newblock {\em Journal of Physics: Conference Series}, 1827(1):012066, March 2021.

\bibitem{manishaGenerativeAdversarialNetworks2018}
P~Manisha and Sujit Gujar.
\newblock Generative {{Adversarial Networks}} ({{GANs}}): {{What}} it can generate and {{What}} it cannot?, 2018.

\bibitem{shurrabSelfsupervisedLearningMethods2022}
Saeed Shurrab and Rehab Duwairi.
\newblock Self-supervised learning methods and applications in medical imaging analysis: A survey.
\newblock {\em PeerJ Computer Science}, 8:e1045, July 2022.

\bibitem{huangSelfsupervisedLearningMedical2023}
Shih-Cheng Huang, Anuj Pareek, Malte Jensen, Matthew~P. Lungren, Serena Yeung, and Akshay~S. Chaudhari.
\newblock Self-supervised learning for medical image classification: A systematic review and implementation guidelines.
\newblock {\em npj Digital Medicine}, 6(1):74, April 2023.

\bibitem{chenSelfsupervisedLearningMedical2019}
Liang Chen, Paul Bentley, Kensaku Mori, Kazunari Misawa, Michitaka Fujiwara, and Daniel Rueckert.
\newblock Self-supervised learning for medical image analysis using image context restoration.
\newblock {\em Medical image analysis}, 58:101539, December 2019.

\bibitem{tajbakhshSurrogateSupervisionMedical2019}
Nima Tajbakhsh, Yufei Hu, Junli Cao, Xingjian Yan, Yi~Xiao, Yong Lu, Jianming Liang, Demetri Terzopoulos, and Xiaowei Ding.
\newblock Surrogate {{Supervision}} for {{Medical Image Analysis}}: {{Effective Deep Learning From Limited Quantities}} of {{Labeled Data}}.
\newblock In {\em 2019 {{IEEE}} 16th {{International Symposium}} on {{Biomedical Imaging}} ({{ISBI}} 2019)}, pages 1251--1255, Venice, Italy, April 2019. IEEE.

\bibitem{peroneUnsupervisedDomainAdaptation2019}
Christian~S. Perone, Pedro Ballester, Rodrigo~C. Barros, and Julien {Cohen-Adad}.
\newblock Unsupervised domain adaptation for medical imaging segmentation with self-ensembling.
\newblock {\em NeuroImage}, 194:1--11, July 2019.

\bibitem{kumariDeepLearningUnsupervised2024}
Suruchi Kumari and Pravendra Singh.
\newblock Deep learning for unsupervised domain adaptation in medical imaging: {{Recent}} advancements and future perspectives.
\newblock {\em Computers in Biology and Medicine}, 170:107912, March 2024.

\bibitem{kimTransferLearningMedical2022}
Hee~E. Kim, Alejandro {Cosa-Linan}, Nandhini Santhanam, Mahboubeh Jannesari, Mate~E. Maros, and Thomas Ganslandt.
\newblock Transfer learning for medical image classification: A literature review.
\newblock {\em BMC Medical Imaging}, 22(1):69, April 2022.

\bibitem{karimiTransferLearningMedical2021}
Davood Karimi, Simon~K. Warfield, and Ali Gholipour.
\newblock Transfer {{Learning}} in {{Medical Image Segmentation}}: {{New Insights}} from {{Analysis}} of the {{Dynamics}} of {{Model Parameters}} and {{Learned Representations}}.
\newblock {\em Artificial intelligence in medicine}, 116:102078, June 2021.

\bibitem{guanDomainAdaptationMedical2022}
Hao Guan and Mingxia Liu.
\newblock Domain {{Adaptation}} for {{Medical Image Analysis}}: {{A Survey}}.
\newblock {\em IEEE transactions on bio-medical engineering}, 69(3):1173--1185, March 2022.

\bibitem{kanakasabapathyAdaptiveAdversarialNeural2021}
Manoj~Kumar Kanakasabapathy, Prudhvi Thirumalaraju, Hemanth Kandula, Fenil Doshi, Anjali~Devi Sivakumar, Deeksha Kartik, Raghav Gupta, Rohan Pooniwala, John~A. Branda, Athe~M. Tsibris, Daniel~R. Kuritzkes, John~C. Petrozza, Charles~L. Bormann, and Hadi Shafiee.
\newblock Adaptive adversarial neural networks for the analysis of lossy and domain-shifted datasets of medical images.
\newblock {\em Nature biomedical engineering}, 5(6):571--585, June 2021.

\bibitem{tangGeneralizableFrameworkUnpaired2022}
Kunning Tang, Ying Da~Wang, James McClure, Cheng Chen, Peyman Mostaghimi, and Ryan~T. Armstrong.
\newblock Generalizable {{Framework}} of {{Unpaired Domain Transfer}} and {{Deep Learning}} for the {{Processing}} of {{Real-Time Synchrotron-Based X-Ray Microcomputed Tomography Images}} of {{Complex Structures}}.
\newblock {\em Physical Review Applied}, 17(3):034048, March 2022.

\bibitem{manchesterDataWorsener2025}
Tristan Manchester.
\newblock Data {{Worsener}}.
\newblock https://github.com/tristanmanchester/data\_worsener, 2025.

\bibitem{xieSegFormerSimpleEfficient2021}
Enze Xie, Wenhai Wang, Zhiding Yu, Anima Anandkumar, Jose~M. Alvarez, and Ping Luo.
\newblock {{SegFormer}}: Simple and efficient design for semantic segmentation with transformers.
\newblock In {\em Proceedings of the 35th International Conference on Neural Information Processing Systems}, Nips '21, Red Hook, NY, USA, 2021. Curran Associates Inc.

\bibitem{SegFormerHuggingFace}
{{SegFormer}} - {{Hugging Face}}.
\newblock https://huggingface.co/docs/transformers/en/model\_doc/segformer.

\bibitem{buslaevAlbumentationsFastFlexible2020}
Alexander Buslaev, Vladimir~I. Iglovikov, Eugene Khvedchenya, Alex Parinov, Mikhail Druzhinin, and Alexandr~A. Kalinin.
\newblock Albumentations: {{Fast}} and {{Flexible Image Augmentations}}.
\newblock {\em Information}, 11(2):125, February 2020.

\bibitem{bermanLovaszSoftmaxLossTractable2018}
Maxim Berman, Amal~Rannen Triki, and Matthew~B. Blaschko.
\newblock The {{Lovasz-Softmax Loss}}: {{A Tractable Surrogate}} for the {{Optimization}} of the {{Intersection-Over-Union Measure}} in {{Neural Networks}}.
\newblock In {\em 2018 {{IEEE}}/{{CVF Conference}} on {{Computer Vision}} and {{Pattern Recognition}}}, pages 4413--4421, Salt Lake City, UT, June 2018. IEEE.

\bibitem{linFocalLossDense2020}
Tsung-Yi Lin, Priya Goyal, Ross Girshick, Kaiming He, and Piotr Dollar.
\newblock Focal {{Loss}} for {{Dense Object Detection}}.
\newblock {\em IEEE Transactions on Pattern Analysis and Machine Intelligence}, 42(2):318--327, February 2020.

\bibitem{salehiTverskyLossFunction2017}
Seyed Sadegh~Mohseni Salehi, Deniz Erdogmus, and Ali Gholipour.
\newblock Tversky loss function for image segmentation using {{3D}} fully convolutional deep networks.
\newblock In Qian Wang, Yinghuan Shi, Heung-Il Suk, and Kenji Suzuki, editors, {\em Machine Learning in Medical Imaging}, pages 379--387, Cham, 2017. Springer International Publishing.

\bibitem{manchesterScrambledSeg2025}
Tristan Manchester.
\newblock {{ScrambledSeg}}.
\newblock https://github.com/tristanmanchester/ScrambledSeg, 2025.

\end{thebibliography}

\end{document}